\documentclass[two
column,secnumarabic,amssymb, nobibnotes, aps, prd]{revtex4-1}

\usepackage{amssymb,amsmath,amsfonts}
\usepackage{bbm,wasysym}
\usepackage{graphicx}
\usepackage{slashbox}
\usepackage{color}
\usepackage{verbatim}
\newcommand{\id}{{\mathbbm{1}}}

\newcommand{\bra}[1]{{\langle #1 |}}
\newcommand{\ket}[1]{{| #1 \rangle}}

\newcommand{\tr}{tr}

\begin{document}

\title{Distributed super dense coding over noisy channels}

\author{Z. Shadman }
\email[]{shadman@thphy.uni-duesseldorf.de}
\author{ H. Kampermann }{}
\author{D. Bru\ss}
\affiliation{ Institute f\"ur Theoretische Physik III,
              Heinrich-Heine-Universit\"at
              D\"usseldorf, D-40225 D\"usseldorf, Germany}
\author{C. Macchiavello}
\affiliation{ Dipartimento di Fisica  ``A. Volta" and INFN-Sezione di Pavia, 
Via Bassi 6, 27100, Pavia, Italy}

\date{\today}%

\begin{abstract}
We study  multipartite super dense coding in the presence of a covariant 
noisy channel. We investigate  the case of many senders and one receiver, considering both unitary and non-unitary encoding. We study the scenarios where the senders apply local encoding or global encoding.  We
 show that, up to some pre-processing on the original state,  the senders  cannot do better encoding than local, unitary encoding.  We then introduce general 
Pauli channels as a significant example of  covariant maps. Considering Pauli channels, we   provide examples for which the super dense coding capacity is explicitly determined.

 \begin{description}{}
\item[PACS numbers] 03.67.-a, 03.67.Hk, 03.65.Ud
\end{description}
\end{abstract}
\maketitle

\section{Introduction}

The notion of multipartite super dense coding was introduced
by Bose \emph{et al.} \cite{Bose-multi-first} to generalize
the Bennett-Wiesner scheme \cite{Bennett} of super dense coding to
multiparties. In this scheme it was shown that the use of a multipartite entangled state can allow a single receiver to read messages from more than one source through a single measurement.  A generalization of this
multipartite super dense coding to higher dimensions was
given by Liu \emph{et al.} \cite{multi-sdc-Liu-higher-dim}. Distributed super dense coding was  also widely discussed in 
\cite{ourPRL,Dagmar} in which two scenarios of many senders with
either one or two receiver(s) were addressed.  For a single receiver, the exact super dense coding capacity was determined and it was shown that the senders do not need to apply
global unitaries to reach the optimal capacity,
but each sender can perform a local encoding on her
side. As a result, it was shown that   bound entangled states with respect to a 
 bipartite cut between the senders (Alices) and the receiver (Bob) are not ``multi''
dense-codeable. Furthermore, a general
classification of multipartite quantum states according to their
dense-codeability was investigated. 

The above multipartite scenarios were discussed for noiseless systems. 
However, in a realistic super dense coding scheme noise is unavoidably 
present in the system. We assume here that noise is  present only  in the transmission channels and the other apparatuses involved are perfect. 
In \cite{zahra-paper,zahra-paper2}, the  bipartite super dense coding for both correlated and uncorrelated channels was discussed.  In the present paper
we generalize those schemes to the  multipartite case in the presence of covariant noise. We investigate the scenario of more than one sender  with a single 
receiver,  considering both unitary and non-unitary encoding. 
We follow two avenues. First, we will consider the case where the senders 
are far apart and can only apply local operations. Second, we will assume that
the senders are allowed to perform global operations.   Since the amount of classical information that can be extracted from an ensemble of quantum states  can be  measured by the Holevo quantity \cite{Gordon, Levitin, Holevo-chi-quantity},  the super dense coding capacity for a given resource  state 
is defined to be the maximal amount of this quantity  with respect to the 
encoding procedure. In the present paper we  focus on the optimization 
problem of the Holevo quantity in order to find the super dense coding 
capacity, considering local (non)unitary encoding  as well as  global  
(non)unitary encoding. 

The paper is organized as follows.
In   Sec. II  we first review the mathematical definition of the Holevo quantity  for an ensemble of multipartite states when the parties are 
connected through a completely positive trace preserving map (a noisy channel). Considering unitary encoding,  and a covariant channel,  for  both scenarios of local and global encoding,  we then find  an expression for the super dense coding capacity. This 
expression only involves to find a  single unitary operator acting on the resource state. In  Sec. III  we discuss the Pauli channel as a  typical example of a covariant map. We then give   examples of  Pauli channels and initial states for which the single unitary operator is explicitly determined. 
In Sec. IV, considering non-unitary encoding,  we derive the  multipartite 
super dense  coding capacity in the presence of  covariant channels up to a pre-processing on the resource state. 
We investigate both local and global encoding. 
Furthermore,  we discuss the Pauli channel as particular map.  In Sec. V 
we summarize the main results. Finally, in the Appendix, we provide  proofs for 
 two Lemmas reported in the paper. 
\section {super dense coding  capacity with many senders and one receiver in the presence of noisy channels\label{sec2}}

A quantum channel is a communication channel
which can transmit a quantum system and can be used to carry classical information. If the transfer is undisturbed the channel is noiseless; if the quantum 
system interacts with some other external systems (environment), a noisy quantum channel results.  
Mathematically,  a quantum channel can be described  as a completely positive trace
preserving (CPTP) map acting  on the quantum state  that is transmitted. Considering a noisy transmission channel, the multipartite super dense coding scheme  works as follows:  
a given quantum state $\rho^{\textmd {a}_1...\textmd {a}_k \textmd {b}}$   is distributed between $k$ Alices and a single Bob (in our scenario,   Bob's subsystem experiences  noise in this stage). Then,  Alices  perform with the  probability $p_{\{i\}}$ a unitary operation  $W_{\{i\}}^{\textmd {a}_1...\textmd {a}_k}$ on their side of the state $\rho^{\textmd {a}_1...\textmd {a}_k\textmd {b}}$,  thus  encoding  classical information through the state  $\rho_ {\{i\}}=(W_ {\{i\}}^{\textmd {a}_1...\textmd {a}_k} \otimes \id^b) \hspace{2mm} \rho^{\textmd {a}_1...\textmd {a}_k\textmd {b}} \hspace{2mm} (W_{\{i\}} ^{ \textmd {a}_1,...,\textmd {a}_k\dagger} \otimes \id^\textmd {b})$, where $\id^\textmd {b}$  is the identity operator on the  Bob's Hilbert space and ${\{i\}}$, is a set of indices for Alices. Subsequently,  the Alices send their subsystems of the encoded state  through the noisy channel to Bob. We consider
$\Lambda_{\textmd {a}_1...\textmd {a}_k\textmd {b}}:\rho_  {\{i\}}\rightarrow \Lambda_{\textmd {a}_1...\textmd {a}_k\textmd {b}}(\rho_ {\{i\}})$ to be the CPTP map (quantum channel) that
globally acts on the multipartite  encoded  state $\rho_  {\{i\}}$. By this process, Bob receives the ensemble $\{\Lambda_{\textmd {a}_1...\textmd {a}_k\textmd {b}}(\rho_ {\{i\}}) , p_{\{i\}}\}$. By performing suitable measurements,  Bob can  extract the  accessible  information about  this ensemble which is  given by the Holevo quantity \cite{ Holevo-chi-quantity}

\begin{eqnarray}
\chi_{\textmd {un}} \left(\{\rho_{\{i\}},p_{\{i\}}\}\right)&=&S\Big (
\sum_{\{i\}} p_{\{i\}}
 \Lambda_{\textmd {a}_1...\textmd {a}_k\textmd {b}}\left(\rho_{\{i\}}  \right)\Big)\nonumber\\
&-&\sum_{\{i\}} p_{\{i\}} S\left(\Lambda_{\textmd {a}_1...\textmd {a}_k\textmd {b}} \left(
\rho_{\{i\}}\right)\right),
 \label{Holevo-multi}
\end{eqnarray}
where $S(\eta)=-\mathrm{\tr}(\eta \log \eta) $ is the von Neumann entropy, and the  logarithm is taken to base two. The subscript \textquotedblleft \textmd{un}\textquotedblright \hspace{0.2mm} refers to unitary encoding. The super dense coding capacity
$C_{\textmd {un}}$ for a given resource state $\rho^{\textmd {a}_1...\textmd {a}_k\textmd {b}}$ and the noisy channel
$\Lambda_{\textmd {a}_1...\textmd {a}_k\textmd {b}}$ is defined to be the maximum of the  Holevo quantity
$ \chi _{\textmd {un}}\left(\{\rho_{\{i\}},p_{\{i\}}\}\right)$  with respect to  the encoding
$\{W_{\{i\}}^{\textmd {a}_1...\textmd {a}_k} , p_{\{i\}}\}$, i.e.

\begin{eqnarray}
&&C_{\textmd{un}}=\max_{\{W_{\{i\}}^{\textmd {a}_1...\textmd {a}_k} , p_{\{i\}}\}}\chi_{\textmd{un}} \left(\{{\rho_{\{i\}}},p_{\{i\}}\}\right).
\label{c-multi-optimization}
\end{eqnarray}
For an illustration, see  Fig. 1. 
\newpage

\begin{center}
\begin{tabular}{c  }
\includegraphics[width=9cm]{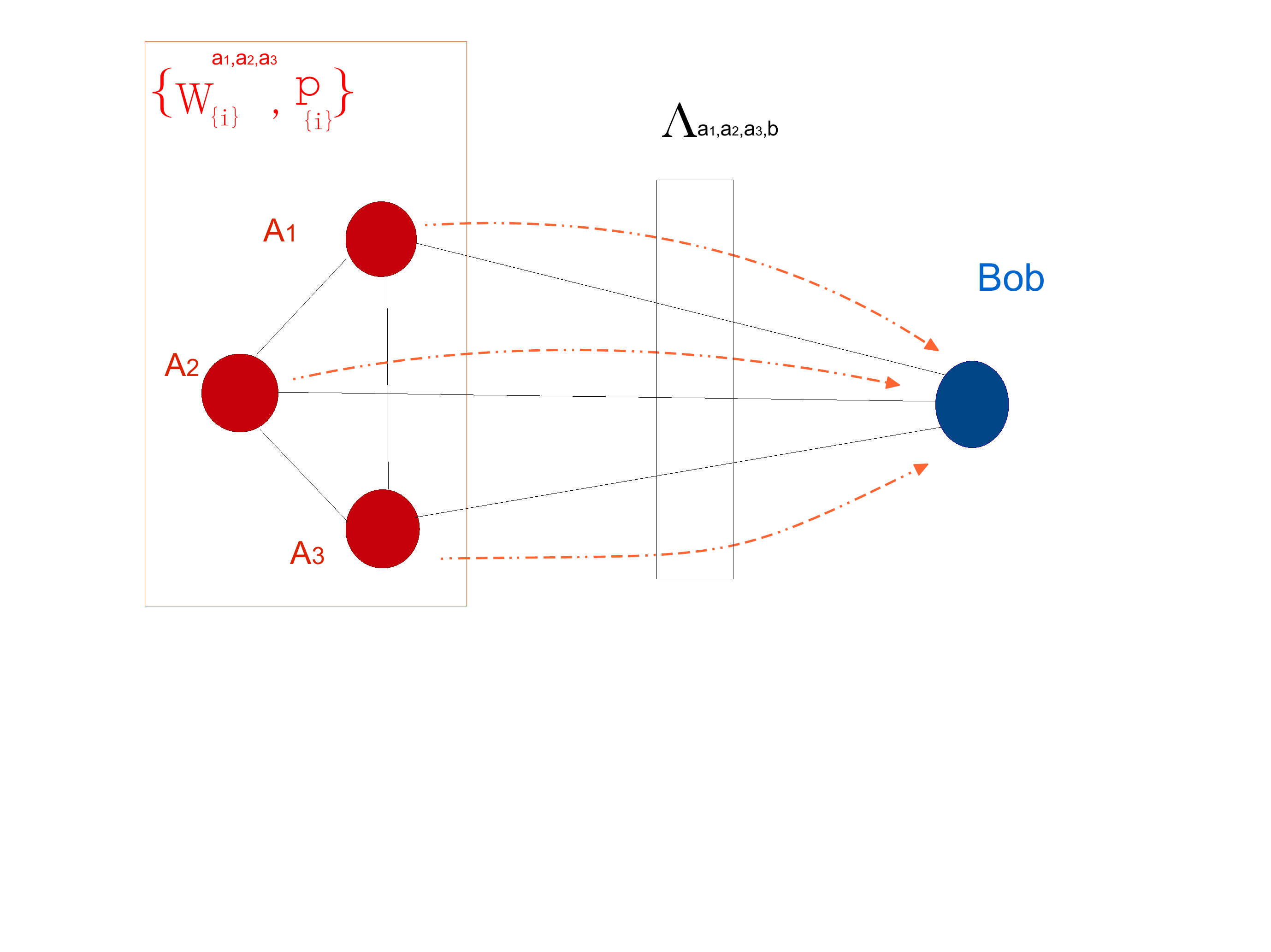}
\end{tabular}
\label{multi}
\end{center}
Fig. 1. (Color online)   Super dense coding with a distributed quantum state $\rho^{\textmd {a}_1 \textmd {a}_2 \textmd {a}_3\textmd {b}}$ between  four parties (three Alices and a single Bob).  The straight black lines show the entanglement between  the  parties and the  dashed red curves show the transmission channels between Alices and Bob.  The  Alices encode with the ensemble  $\{ W_{\{i\}}^{\textmd{a}_1\textmd {a}_2\textmd{a}_3}, p_{\{i\}}\}$ and in the next step, they send  their subsystems of the encoded state  through the channel  $\Lambda_{\textmd {a}_1\textmd {a}_2\textmd {a}_k\textmd {b}}$ to the receiver, Bob. The ensemble that Bob gets is $\{\Lambda_{\textmd {a}_1\textmd {a}_2\textmd {a}_k\textmd {b}}(\rho_ {\{i\}}), p_{\{i\}} \}$. In this process, based on  optimal encoding by  the Alices,  the maximal amount of  classical information which is defined to be the capacity is transferred   (see main text). 


\subsection{ Covariant noisy channels}

In this section we determine the super dense coding capacity for a special class of channels, up to a single unitary operator  acting on  the given state $\rho^{\textmd {a}_1...\textmd {a}_k\textmd {b}} $.  
The channels we consider, denoted by  
$\Lambda_{\textmd{a}_1...\textmd{a}_k\textmd{b}}^\textmd {c }$,  
commute with a complete set of orthogonal unitary operators 
$ \tilde{V}_{\{i\}}$, namely they have   the property 

\begin{eqnarray}
\Lambda^\textmd{c}_{\textmd{a}_1...\textmd{a}_k\textmd{b}}(  \tilde{V}_{\{i\}} \rho   \tilde{V}_{\{i\}}^\dagger) = 
  \tilde{V}_{\{i\}} \Lambda^\textmd{c}_{\textmd{a}_1...\textmd{a}_k\textmd{b}}(\rho)   \tilde{V}_{\{i\}}^\dagger, 
\label{covariance}
\end{eqnarray}
for the set of unitary operators which satisfy the orthogonality 
condition ${\mathrm{\tr}}[\tilde{V}_i\tilde{V}_j^\dagger]=d \delta_{ij}$.
According to \cite{hiroshima}, for this set it is guaranteed that 
$ \frac{1}{d}\sum_i U_i \Xi {U_i^\dagger} = \id \mathrm{\mathrm{\tr}} \Xi$ where $ \Xi$ is an arbitrary operator.
The property (\ref{covariance}) is usually referred to as covariance
\cite{holevo}.
Here we will consider local unitary operators, namely of the form 
   \begin{eqnarray}
  \tilde{V}_{\{i\}}=V_{i_1}^{\textmd{a}_1}\otimes...\otimes V_{i_k}^{\textmd{a}_k}.
  \label{Vtilde}
  \end{eqnarray}

In the following  we will first  discuss the case that  the $k$ Alices are far 
apart and  they are restricted to local unitary operations in the presence of a covariant channel with the property (\ref{covariance}). 
We will then investigate the case  where the Alices are allowed to  perform 
entangled unitary encoding.

\subsubsection {Senders performing local unitary operators \label{unitary}}

In this scenario, the $j$th  Alice applies  a local unitary operator 
$W_{i_j}^{\textmd {a}_j}$  
with probability $p_{i_j}$ on her subsystem of the shared state $\rho^{\textmd {a}_1...\textmd {a}_k\textmd {b}}$.  The optimization of 
the Holevo quantity  is given in the  following lemma. \\

\textbf{Lemma 1.} Let
\begin{eqnarray}
\chi_ {\textmd {un}}^{\textmd {lo}}&=&S\Big (
\sum_{\{i\}} p_{\{i\}}
 \Lambda_{\textmd {a}_1...\textmd {a}_k\textmd {b}}^\textmd {c}\left(\rho_{\{i\}}  \right)\Big)\nonumber\\
&-&\sum_{\{i\}} p_{\{i\}} S\left(\Lambda_{\textmd {a}_1...\textmd {a}_k\textmd {b}}^\textmd {c} \left(
\rho_{\{i\}}\right)\right),
 \label{Holevo-multi-pauli}
\end{eqnarray}
be the Holevo quantity with  

\begin{eqnarray}
\rho_ {\{i\}}&=& ( W_{i_1}^{\textmd {a}_1}\otimes W_{i_2}^{a_2}\otimes...\otimes W_{i_k}^{\textmd {a}_k}   \otimes \id^\textmd {b}) \hspace{2mm} \rho^{\textmd {a}_1...\textmd {a}_k\textmd {b}}\nonumber\\
&& \hspace{2mm} (W_{i_1}^{\textmd {a}_1}\otimes W_{i_2}^{\textmd {a}_2 \dagger}\otimes...\otimes W_{i_k}^{\textmd {a}_k} \otimes \id^ \textmd {b}),
\end{eqnarray}
 and $\Lambda_{\textmd {a}_1...\textmd {a}_k\textmd {b}}^\textmd {c}$  be a
covariant channel with the property 
(\ref{covariance}).  The superscript  \textquotedblleft \textmd{lo}\textquotedblright \hspace{0.2mm} refers to local encoding. Let
\begin{eqnarray}
U_{\textmd{min}}^{\textmd{lo}}:=U_\textmd{min}^{\textmd{a}_1}  \otimes... \otimes U_\textmd{min}^{\textmd{a}_k} 
\end{eqnarray}
  be the unitary operator
that minimizes the von Neumann entropy after application of this
unitary operator and  the channel  $\Lambda_{\textmd {a}_1...\textmd {a}_k\textmd {b}}^\textmd {c}$  to the initial
state $\rho^{\textmd {a}_1...\textmd {a}_k\textmd {b}} $, i.e.  $U_{\textmd{min}}^{\textmd{lo}}$ minimizes the expression
 $S\left(\Lambda_{\textmd{a}_1...\textmd{a}_k\textmd{b}}^\textmd{c}  \Big( \big(U_{\textmd{min}}^{\textmd{lo}}
 \otimes \id^\textmd{b} \big)   \rho^{\textmd{a}_1...\textmd{a}_k\textmd{b}} \hspace{1mm} \big(   U_{\textmd{min}}^{\textmd{lo}\dagger} \otimes\id^\textmd{b} \big)\Big)\right)$.
 Then the super dense coding capacity  $C_{\textmd{un}}^{\textmd{lo}}$ is given by
\begin{eqnarray}
&C&_{\textmd{un}}^{\textmd{lo}}=\log D_\textmd{A}+ S\left(\Lambda_{\textmd{b}}^\textmd{} \left(
\rho_{\textmd{b}}\right)\right)\nonumber\\
&&- S\bigg(\Lambda_{\textmd{a}_1...\textmd{a}_k\textmd{b}}^\textmd{c} \Big(\big(U_{\textmd{min}}^{\textmd{lo}}
 \otimes \id^\textmd{b} \big)  \rho^{\textmd{a}_1...\textmd{a}_k\textmd{b}} \hspace{1mm} \big(   U_{\textmd{min}}^{\textmd{lo}\dagger} \otimes\id^\textmd{b} \big)\Big)\bigg), \nonumber\\
\label{local-dc-covariant}
\end{eqnarray}
where  $D_\textmd{A}=d_{\textmd{a}_1} d_{\textmd{a}_2}...d_{\textmd{a}_k}$ is the dimension of the Hilbert space of  the $k$ Alices,  and $\mathrm{\tr}_{\textmd{a}_1...\textmd{a}_k}  \Lambda_{\textmd{a}_1...\textmd{a}_k\textmd{b}}^\textmd{c} \left(\rho^ {\textmd{a}_1...\textmd{a}_k\textmd{b}} \right)=\Lambda_{\mathrm{b}}^\textmd{} \left(
\rho_{\mathrm{b}}\right)$. \\
\textbf{Proof:} The von Neumann entropy is subadditive. The maximum entropy of
a $D_A$-dimensional system is $\log D_A $. Since  $ U_{\textmd{min}}^{\textmd{lo}} $ is a unitary
operator that leads to the minimum of the output  von Neumann
entropy, an upper bound on Holevo quantity (\ref{Holevo-multi-pauli})  can be given  as
\begin{eqnarray}
\chi_ {\textmd {un}}^{\textmd {lo}}&\leq& S\Big(\sum_{\{i\}} p_{\{i\}}
 \Lambda_{\textmd{a}_1...\textmd{a}_k\textmd{b}}^\textmd{c} \left(\rho_{\{i\}}  \right)\Big)- S\bigg(\Lambda_{\textmd{a}_1...\textmd{a}_k\textmd{b}}^\textmd{c}  \nonumber\\
&& \Big(\big( U_{\textmd{min}}^{\textmd{lo}} \otimes\id^\textmd{b}\big) \rho^{\textmd{a}_1...\textmd{a}_k\textmd{b}} \hspace{1mm} \big(   U_{\textmd{min}}^{\textmd{lo}\dagger} \otimes\id^\textmd{b} \big)\Big)\bigg) \nonumber\\
&\leq& \log D_\textmd{A}+ S\left(\Lambda_{\textmd{b}}^\textmd{} 
\rho_{\textmd{b}}\right) - S\bigg(\Lambda_{\textmd{a}_1...\textmd{a}_k\textmd{b}}^\textmd{c} \Big(\big(  U_{\textmd{min}}^{\textmd{lo}}  \otimes\id^\textmd{b}\big) \left.   \right. \nonumber\\
&& \rho^{\textmd{a}_1...\textmd{a}_k\textmd{b}} \hspace{1mm} \big(   U_{\textmd{min}}^{\textmd{lo}\dagger} \otimes\id^\textmd{b} \big) \Big)\bigg).
\label{upper-bound-multi}
\end{eqnarray}
In the next step, we show that the upper bound (\ref{upper-bound-multi}) is reachable by the ensemble
$\{\tilde{U}_{\{i\}}=   \tilde{V}_{\{i\}} U_{\textmd{min}}^{\textmd{lo}},\tilde{p}_{\{i\}}=\frac{1}{D^2_{\textmd{A}}} \}$
 where  $\tilde{V}_{\{i\}} $ was defined in Eqs. (\ref{covariance}) and (\ref{Vtilde}). 
 
 The Holevo quantity for
the ensemble $\{\tilde{U}_{\{i\}}, \tilde{p}_{\{i\}} \}$ is denoted by $\tilde{\chi}_{\textmd {un}}^{\textmd {lo}}$ and is given by
\begin{eqnarray}
\tilde{\chi}_{\textmd {un}}^{\textmd {lo}}&=& S \bigg( \sum_{\{i\}}\frac{1}{D_\textmd A^2}\Lambda^{\textmd c}_{\textmd{a}_1...\textmd{a}_k\textmd{b}}\left( \tilde{U}_{\{i\}}\rho^{\textmd{a}_1...\textmd{a}_k\textmd{b}} {\tilde{U}_{\{i\}}}^{\dagger}
 \right)\bigg)\nonumber\\
&-&\sum_{\{i\}} \frac{1}{D_\textmd A^2}
S\bigg(\Lambda^{\textmd c}_{\textmd{a}_1...\textmd{a}_k\textmd{b}}\left(
\tilde{U}_{\{i\}}\rho^{\textmd{a}_1...\textmd{a}_k\textmd{b}}{\tilde{U}_{\{i\}}}^{\dagger} \right)\bigg).
\label{optiensemble-unitary}
\end{eqnarray}
  By using the covariance property (\ref{covariance}), the argument in
the first term on the RHS of (\ref{optiensemble-unitary}) is given
by
\begin{eqnarray}
&& \sum_{\{i\}}\frac{1}{D_\textmd A^2}\Lambda^{\textmd c}_{\textmd{a}_1...\textmd{a}_k\textmd{b}}\left(
(\tilde{U}_{\{i\}}\otimes\id^{\textmd b})\rho^{\textmd{a}_1...\textmd{a}_k\textmd{b}}({\tilde{U}_{\{i\}}}^{\dagger}\otimes\id^{\textmd b})
 \right)\nonumber\\
 &=& \frac{1}{D_\textmd A^2} \sum_{{\{i\}}} (\tilde{V}_{{\{i\}}}\otimes \id^\textmd{b} ) \underbrace{
\Big[\Lambda_{\textmd{a}_1...\textmd{a}_k\textmd{b}}^\textmd{c} \Big(\big(U_{\textmd{min}}^{\textmd{lo}} \otimes    \id^\textmd{b}\big)\rho^{\textmd{a}_1...\textmd{a}_k\textmd{b}}  }\nonumber\\
&& \underbrace{ \left.  \big(U_{\textmd{min}}^{\textmd{lo }\dagger}\otimes\id^\textmd{b} \big)\right)\Big]  }_{:= \varrho}
(\tilde{V}_{{\{i\}}}^\dagger\otimes \id^\textmd{b})
\end{eqnarray}
The density matrix  $\varrho$  with the bipartite cut between the Alices and Bob, and in the Hilbert-Schmidt representation, can be  decomposed as
\begin{eqnarray}
\varrho &=&\frac{\id^{\textmd{a}_1...\textmd{a}_k}}{D_\textmd{A}}\otimes\Lambda _{\textmd{b}}(\rho_\textmd{b})+\sum_j r_j\lambda_j^{\textmd{a}_1...\textmd{a}_k}\otimes\id^{\textmd{b}}\nonumber\\
&+&\sum_{j,k}t_{jk} \lambda_j^{\textmd{a}_1...\textmd{a}_k}\otimes\lambda_k^\textmd{b} \hspace{1mm} ,
\label{decomposition-multi}
\end{eqnarray}
where the  $\lambda_j^{\textmd{a}_1...\textmd{a}_k}$ are the generators of	the	$SU(D_\textmd A)$	algebra,  and $\lambda_k ^\textmd b$  are the generators of the $SU(d_\textmd b)$ algebra with $\textmd{\tr}\lambda_j=0$. The	parameters $r_j$ and $t_{jk}$ are real numbers. 
By exploiting  the equation 
$ \frac{1}{d}\sum_i \tilde{V}_{\{i\}} \Xi \tilde{V}_{\{i\}} ^\dagger = 
\id \mathrm{\mathrm{\tr}} \Xi$, and since each $\lambda_j$ is traceless, 
we can 
write 
\begin{eqnarray}
 \sum_{\{i\}} 
\tilde{V}_{\{i\}} \lambda_j^{\textmd{a}_1...\textmd{a}_k}  {\tilde{V}}_{\{i\}}^{\dagger}=0.
\label{average-lambda}
\end{eqnarray}
By using this property and the decomposition (\ref{decomposition-multi}),  
we find that the argument in
the first term on the RHS of (\ref{optiensemble-unitary}) is given
by
\begin{eqnarray}
&& S \bigg( \sum_{\{i\}}\frac{1}{D_\textmd A^2}\Lambda^{\textmd c}_{\textmd{a}_1...\textmd{a}_k\textmd{b}}\left( \tilde{U}_{\{i\}}\rho^{\textmd{a}_1...\textmd{a}_k\textmd{b}} {\tilde{U}_{\{i\}}}^{\dagger}
 \right)\bigg) \nonumber\\
&&=\log D_\textmd{A}+ S\left(\Lambda_{\textmd{b}}^\textmd{} \left(
\rho_{\textmd{b}}\right)\right).
\label{average-state1}
\end{eqnarray}
Furthermore, the second term on the RHS of Eq. (\ref{optiensemble-unitary})  can be expressed in terms of the
unitary operator $U_{\textmd{min}}^{\textmd{lo}}$ and the channel. By using the covariance property (\ref{covariance}),  and since the von Neumann entropy is invariant under a unitary
transformation, we can write
\begin{eqnarray}
&&\sum_{\{i\}} \frac{1}{D_\textmd A^2}
S\bigg(\Lambda^{\textmd c}_{\textmd{a}_1...\textmd{a}_k\textmd{b}}\left(
(\tilde{U}_{\{i\}}\otimes\id^\textmd{b})\rho^{\textmd{a}_1...\textmd{a}_k\textmd{b}}({\tilde{U}_{\{i\}}}^{\dagger}\otimes\id^\textmd{b}) \right)\bigg)\nonumber\\
&=&\frac{1}{D_\textmd A^2} \sum_{{\{i\}}} S\bigg((\tilde{V}_{{\{i\}}}\otimes \id^\textmd{b} )
\Big[\Lambda_{\textmd{a}_1...\textmd{a}_k\textmd{b}}^\textmd{c} \Big(\big(U_{\textmd{min}}^{\textmd{lo}} \otimes    \id^\textmd{b}\big)\rho^{\textmd{a}_1...\textmd{a}_k\textmd{b}}  \nonumber\\
&&\left.  \big(U_{\textmd{min}}^{\textmd{lo }\dagger}\otimes\id^\textmd{b} \big)\right)\Big]
(\tilde{V}_{{\{i\}}}^\dagger\otimes \id^\textmd{b})\bigg)\nonumber\\
&=& S\Big(\Lambda_{\textmd{a}_1...\textmd{a}_k\textmd{b}}^\textmd{c}  \Big(\big(U_{\textmd{min}}^{\textmd{lo}}
 \otimes \id^\textmd{b} \big) 
\left.   \rho^{\textmd{a}_1...\textmd{a}_k\textmd{b}} \hspace{1mm} \big(   U_{\textmd{min}}^{\textmd{lo}\dagger} \otimes\id^\textmd{b} \big) \Big)\right).\nonumber\\
\label{average-entropy1}
\end{eqnarray}
 Inserting Eqs. (\ref{average-state1}) and (\ref{average-entropy1}) into
Eq. (\ref{optiensemble-unitary}), one finds that the Holevo quantity
$\tilde{\chi}_{\textmd {un}}^{\textmd {lo}} $ is equal to  the upper bound given in
Eq. (\ref{upper-bound-multi}) and therefore this is the super dense
coding capacity. \hfill {$\Box$}

As we can see from the capacity expression (\ref{local-dc-covariant}),  all the parameters are known except the single unitary operator $ U_{\textmd{min}}^{\textmd{lo}}$.  However, for some specific situations like  noiseless channels, 
i.e. for $ \Lambda^{\textmd c}_{\textmd{a}_1...\textmd{a}_k\textmd{b}}=\id$, 
 this unitary operator has  already been identified as the identity operator.  
The capacity for noiseless channels is then given by 
 $C=\log D_\textmd{A}+ S(
\rho_{\textmd{b}} ) - S( \rho^{\textmd{a}_1...\textmd{a}_k\textmd{b}})$. We also provide more examples in the next section. 
\subsubsection{Senders may perform entangled unitaries}
We will now investigate the case where the Alices are allowed  to apply entangled 
unitary operators. The question we want to address is:  can the Alices  increase the  information transfer by applying entangled unitaries? To answer this 
question we follow a strategy similar to the case of local encoding, mentioned in the previous part.  The difference is that instead of local unitaries $ W_{i_j}^{\textmd {a}_j} $,  Alices encode with the  global unitary operators $W_ {\{i\}}^{\textmd {a}_1...\textmd {a}_k} $ with the probabilities $p_{\{i\}}$. In order to find the optimal encoding and thus the super dense coding capacity, we optimize the Holevo quantity (\ref{Holevo-multi}). The optimization procedure is similar to Lemma 1. The difference is that we now have a global unitary  operator $U_\textmd{min}^{\textmd{g}}$ which  minimizes  the output von Neumann entropy. We  can then show that the optimal encoding is given by the ensemble $\{\tilde{U}_{\{i\}}=   \tilde{V}_{\{i\}} U_\textmd{min}^{\textmd{g}},\tilde{p}_{\{i\}}=\frac{1}{D^2_\textmd{A}} \}$, and the super dense coding capacity $C_{\textmd{un}}^\textmd{g}$ for this situation is given by 
\begin{eqnarray}
&&C_{\textmd{un}}^\textmd{g}=\log D_\textmd{A}+ S\left(\Lambda_{\textmd{b}}^\textmd{} \left(
\rho_{\textmd{b}}\right)\right)  \nonumber\\
&&-S\bigg(\Lambda_{\textmd{a}_1...\textmd{a}_k\textmd{b}}^\textmd{c} \Big( \big(U_\textmd{min}^{\textmd{g}} 
\otimes\id^\textmd{b}\big)\rho^{\textmd{a}_1...\textmd{a}_k\textmd{b}}\big(U_\textmd{min}^{ \textmd{g}\dagger}  \otimes\id^\textmd{b} \big)\Big)\bigg).\nonumber\\
\label{global-dc-covariant}
\end{eqnarray}
The difference between the capacities $(\ref{local-dc-covariant})$ and $(\ref{global-dc-covariant})$ is the occurrence of the local and global unitary transformation $U_{\textmd{min}}^{\textmd{lo}}$ and $U_\textmd{min}^{ \textmd{g}}$, respectively.

\section{ Pauli noise as a model for a covariant quantum channel\label{sec2-1}}
 
In the present section we will consider the explicit case of Pauli channels, 
namely  channels whose action on a $d$-dimensional density 
operator $\xi$ is given by
\begin{eqnarray}
&&\Lambda^{\textmd {P}}(\xi)=\sum_{m,n=0}^{d-1}q_{mn} V_{mn}\xi V_{mn}^\dagger\;,
\label{pauli-d-channel}
\end{eqnarray}
where $V_{mn}$ are the displacement operators defined as
 
\begin{eqnarray}
&&V_{mn}=\sum_{k=0}^{d-1}\exp \left({\frac{2i\pi kn}{d}}\right)\ket{k}\bra{k+m(\mathrm{mod}\,
{d})}\;.
\label{vmn}
\end{eqnarray}
They satisfy $\mathrm{\tr} V_{mn} = d\delta_{m0} \delta_{n0} $, and $V_{mn}
V_{mn}^\dagger=\id$. They also have the properties
\begin{subequations}
\begin{eqnarray}
\label{vvdagger}
&&\textmd {\tr} [V_{mn} V_{{m}^{\prime}{n}^{\prime}}^\dagger] = d\delta_{m{m}^{\prime}} \delta_{n{n}^{\prime}},\\
\label{vv}
&&\nonumber\\
&& V_{mn} V_{{m}^{\prime}{n}^{\prime}}
=\exp\left({\frac{2i\pi({n}^{\prime}m-n{m}^{\prime})}{d}}\right) V_{{m}^{\prime}{n}^{\prime}}V_{mn},\\
\label{groupproduct}
&&V_{mn}V_{{m}^{\prime}{n}^{\prime}}=\exp \left({\frac{2i\pi {n}^{\prime}m }{d}}\right)V_{m+{m}^{\prime}(\textmd {mod}
\, {d}),n+{n}^{\prime}( \textmd {mod}\, {d})}.\nonumber\\
\end{eqnarray}
\end{subequations}

The superscript \textquotedblleft \textmd{P}\textquotedblright \hspace{0.2mm} 
in (\ref{pauli-d-channel}) refers to the Pauli channel. Here  $q_{mn}$ are 
probabilities (i.e. $q_{mn}\geq 0$ and
$\sum_{mn}q_{mn}=1$).  Since the operators $V_{mn} $ are unitary, 
the Pauli channel 
(\ref{pauli-d-channel}) maps the identity to itself (it is a unital channel). 

In the case of $k+1$ parties we can consider a general Pauli channel which
globally acts on the $k$ Alices' subsystems (after encoding) and  Bob's 
subsystem (in the distribution stage) as 
\begin{eqnarray}
&&\Lambda_{\textmd {a}_1...\textmd {a}_k\textmd {b}}^\textmd {P}\big(\xi\big)
=\sum_{\{m_in_i\}}q_{\{m_in_i\}}
 \Big(V_{m_1n_1}^{\textmd {a}_1} \otimes... \otimes V_{m_kn_k}^{\textmd {a}_k}   \otimes \nonumber\\
&&   V_{m_{k+1} n_{k+1}}^{\textmd {b}} \Big)\xi\left({V_{m_1n_1}^{\textmd {a}_1\dagger}}
 \otimes...\otimes {V_{m_kn_k}^{\textmd {a}_k\dagger}}\otimes V_{  m_{k+1} n_{k+1} }^{\textmd {b}\dagger} \right), 
 \label{k+1-channel}
\end{eqnarray}
where the probabilities $q_{\{m_in_i\}}$ add to \emph{one}. Here, the 
notations $\{m_in_i\}_{i=1}^k$  stand for $k$ Alices and ${m_{k+1} n_{k+1}}$ 
stands for Bob. 

Since the displacement operators commute up to a phase,  it is straightforward 
to see that the Pauli channel (\ref{k+1-channel}) is a {covariant} channel. 
Therefore,  the capacities for local and global unitary encoding are 
a special form of Eqs. (\ref{local-dc-covariant}) and 
(\ref{global-dc-covariant}), respectively, and are given by

\begin{eqnarray}
&&C_{\textmd{un}}^\textmd{lo,P}=\log D_\textmd{A}+ S\left(\Lambda_{\textmd{b}}^\textmd{P} \left(
\rho_{\textmd{b}}\right)\right)  \nonumber\\
&&-S\bigg(\Lambda_{\textmd{a}_1...\textmd{a}_k\textmd{b}}^\textmd{c} \Big( \big(U_\textmd{min}^{\textmd{lo}} 
\otimes\id^\textmd{b}\big)\rho^{\textmd{a}_1...\textmd{a}_k\textmd{b}}\big(U_\textmd{min}^{ \textmd{lo}\dagger}  \otimes\id^\textmd{b} \big)\Big)\bigg)
\nonumber\\
\label{local-dc}
\end{eqnarray}
 and 
\begin{eqnarray}
&&C_{\textmd{un}}^\textmd{g,P}=\log D_\textmd{A}+ S\left(\Lambda_{\textmd{b}}^\textmd{P} \left(
\rho_{\textmd{b}}\right)\right)  \nonumber\\
&&-S\bigg(\Lambda_{\textmd{a}_1...\textmd{a}_k\textmd{b}}^\textmd{P} \Big( \big(U_\textmd{min}^{\textmd{g}} 
\otimes\id^\textmd{b}\big)\rho^{\textmd{a}_1...\textmd{a}_k\textmd{b}}\big(U_\textmd{min}^{ \textmd{g}\dagger}  \otimes\id^\textmd{b} \big)\Big)\bigg)\;,
\nonumber\\
\label{global-dc}
\end{eqnarray}
where, in both of the above equations, $\rho_\textmd {b}=\mathrm{\tr}_{\textmd{a}_1...\textmd{a}_k} \rho^{\textmd{a}_1...\textmd{a}_k\textmd{b}}$ represents Bob's reduced density operator and $\Lambda^\textmd{P}_\textmd {b}$ is the $d_\textmd b$-dimensional
Pauli channel (\ref{pauli-d-channel}) acting on  Bob's subsystem. \\

This general model of Pauli channels includes both the case of a memoryless 
channel, where the Pauli noise acts independently on each of the $k+1$ parties
and the probabilities $q_{\{m_in_i\}}$ are products of the single party 
probabilities $q_{mn}$, or more generally the case where the action of noise 
is not independent on consecutive uses but is correlated. 
For example, for $k+1$ uses  of a Pauli channel we can define a  
\emph{correlated} Pauli channel in the multipartite scenario as follows  
\begin{eqnarray}
&&q_{\{m_in_i\}} \nonumber\\
&=&(1-\mu_{12})...(1-\mu_{k ,{k+1}})q_{m_1n_1}...q_{ {m_{k+1} n_{k+1}}}\nonumber\\
&+&\mu_{12}(1-\mu_{13})...(1-\mu_{k,{k+1}})\delta_{m_1m_2}\delta_{n_1n_2}q_{m_1n_1}\nonumber\\
&&q_{m_3n_3}...q_{m_{k+1} n_{k+1}}\nonumber\\
&+&(1-\mu_{12})\mu_{13} ...(1-\mu_{k, {k+1}})\delta_{m_1m_3}\delta_{n_1n_3}q_{m_1n_1}\nonumber\\
&&q_{m_2n_2}q_{m_4n_4}...q_{ m_{k+1} n_{k+1}}\nonumber\\
&.&\nonumber \\
&.&\nonumber \\
&.&\nonumber \\
&+&(1-\mu_{12})...(1-\mu_{k-1 ,{k+1}})\mu_{k, {k+1}}\hspace{0.5mm}\delta_{m_k m_{k+1}}\delta_{n_k n_{k+1}}\nonumber\\
&&q_{m_1n_1}q_{m_3n_3}...q_{m_{k-1}n_{k-1}}q_{m_{k+1}n_{k+1}}\nonumber\\
&+&\mu_{12} \mu_{13}(1-\mu_{14})...(1-\mu_{k, {k+1}})\delta_{m_1m_2}\delta_{n_1n_2}\nonumber\\
&&\delta_{m_1m_3}\delta_{n_1n_3}q_{m_1n_1}q_{m_4n_4}...q_{m_{k+1} n_{k+1}}\nonumber\\
&.&\nonumber \\
&.&\nonumber \\
&.&\nonumber \\
&+&\mu_{12}...\mu_{k-1, {k+1}}(1-\mu_{k,{k+1}})\delta_{m_1m_2}\delta_{n_1n_2}...\delta_{m_1m_k}\delta_{n_1n_k}\nonumber\\
&&q_{m_1n_1}q_{m_{k+1}n_{k+1}}\nonumber\\
&+&\mu_{12}...\mu_{k ,{k+1}}\hspace{0.5mm}\delta_{m_1m_2}\delta_{n_1n_2}...\delta_{m_1m_{k+1}}\delta_{n_1n_{k+1}}q_{m_1n_1}.
\label{qm1n1-mknk}
\end{eqnarray}	
Here, between every two individual channels we have defined a correlation 
degree $\mu_{jl}$  with $0\leq \mu_{jl}\leq 1$  which correlates the channel 
$j$ to the  channel $l$ ($j\neq l$ ).  Thus, for $k+1$ parties we have 
$\frac{k(k+1)}{2}$ correlation degrees $\mu_{jl}$. 
For instance,  $\mu_{12}$  correlates the channel  \emph{one} and   
\emph{two},  $\mu_{k,{k+1}}$ correlates the channel   \emph{k}  and 
Bob's channel, etc. If $\mu_{jl}=0$ for all $j$ and $l$,  then the 
$k+1$ channels are independent or, in other words, we are in the memoryless 
(or uncorrelated) case. As mentioned above, this channel can be expressed 
as a product of independent $k+1$ channels acting seperately  on each subsystem. 
If $\mu_{jl}=1$ for all $j$ and $l$,  we have a \emph{fully correlated} 
Pauli  channel. For other values of  $\mu_{jl}$ other than \emph{zero} and  
\emph{one},  the channel (\ref{k+1-channel}) is partially correlated.  
For two uses of a Pauli channel, the expression (\ref{qm1n1-mknk}) reduces 
to $q_{m_1n_1m_2n_2}=(1-\mu)q_{m_1n_1}q_{m_2n_2}+\mu\delta_{m_1m_2}
\delta_{n_1n_2}q_{m_1n_1}$ with a single correlation degree $\mu$ \cite{mp}. 
We considered this situation in \cite{zahra-paper, zahra-paper2} for the case 
of  bipartite super dense coding.

In the next section, we give examples for which the unitaries  
$U_{\textmd{min}}^{\textmd{lo}}$ and $U_\textmd{min}^{ \textmd{g}}$  are 
determined. For these examples, we show that both  capacities  are the 
same. Thus the  Alices can reach the optimal information transfer via local encoding.

\section{Examples}
In this section,  we show examples of multipartite systems for which 
$U_{\textmd{min}}^{\textmd{lo}}$  or/and $U_{\textmd{min}}^{\textmd{g}}$ are determined. One example is  a \emph{correlated} Pauli channel (\ref{k+1-channel}) and  $k$ copies of the  Bell state. Noise here acts just on the  Alices' subsystem. Another example is  a \emph {fully correlated} Pauli
channel and a GHZ  state  as  well as $k$ copies of a  
Bell  diagonal state, both for $d=2$. The  last example will be the  depolarizing channel with \emph{uncorrelated} noise.

\subsection{$k$ copies of a Bell state and a correlated Pauli channel}

In this section we discuss the example that the Alices and  Bob share $k$ 
copies of the Bell state. We consider the situation when there is no noise on  Bob's side, and the 
Alices' shares of the Bell states  globally experience a \emph{correlated} 
Pauli channel $\Lambda^{\textmd{P}}_{\textmd{a}_1...\textmd{a}_k}$ 
(see Fig. 2). This example satisfies the situation discussed in Sec. 
\ref{unitary}. Therefore, the capacity follows from Eqs. (\ref{local-dc}) and 
(\ref{global-dc}). 

A Bell state in $d\times d$ dimensions is defined as $\ket{\Phi_{00}}= \frac{1}{\sqrt d}\sum_{j=0}^{d-1}\ket{jj}$.
The set of the other maximally entangled Bell states is  denoted by  $\ket{\Phi_{mn}}=(V_{mn}\otimes\id)\ket{\Phi_{00}} $, for $m,n=0,1,...,d-1$. We prove that the von Neumann entropy is 
invariant under arbitrary unitary rotation $ U^{a_1...a_k}$ of the state
$\rho_{00}^{a_1b_1}\otimes...\otimes \rho_{00}^{a_kb_k}$ after application of the channel  $\Lambda^{\textmd{P}}_{\textmd{a}_1...\textmd{a}_k}$, i.e.
\begin{eqnarray}
&&S\bigg(\Lambda^{\textmd{P}}_{\textmd{a}_1...\textmd{a}_k}\Big(\left(U^{\textmd{a}_1...\textmd{a}_k}\otimes\id^{\textmd{b}_1...\textmd{b}_k} \right)
 \left(\rho_{00}^{\textmd{a}_1\textmd{b}_1} \otimes...\otimes \rho_{00}^{\textmd{a}_k\textmd{b}_k} \right)\nonumber\\
&& \left.\big( {U^{\textmd{a}_1...\textmd{a}_k}}^\dagger \otimes\id^{\textmd{b}_1...\textmd{b}_k}
  \big) \right) \bigg)\nonumber\\
&=&S\bigg(\Lambda^{\textmd{P}}_{\textmd{a}_1...\textmd{a}_k}
 \left(\rho_{00}^{\textmd{a}_1\textmd{b}_1}\otimes...\otimes \rho_{00}^{\textmd{a}_k\textmd{b}_k} \right) \bigg).
\end{eqnarray}
To show this claim,  we first prove the following lemma.\\

\textbf{Lemma
2.} Let
\begin{eqnarray}
\rho_{00}^{\textmd{a}_1\textmd{b}_1}\otimes...\otimes \rho_{00}^{\textmd{a}_k\textmd{b}_k}=\ket{\Phi_{00}^{\textmd{a}_1\textmd{b}_1}...\Phi_{00}^{\textmd{a}_k\textmd{b}_k}}
\bra{\Phi_{00}^{\textmd{a}_1\textmd{b}_1}...\Phi_{00}^{\textmd{a}_k\textmd{b}_k}},\nonumber\\
\label{Bell-multi}
\end{eqnarray}
be $k$ copies of the Bell states with different dimensions $d^2_{j}$. Let us define
\begin{eqnarray}
&&\pi_{\{m_in_i\}}:=\left(V_{m_1n_1}^{\textmd{a}_1} \otimes ... \otimes V_{m_kn_k}^{\textmd{a}_k} \otimes\id^{\textmd{b}_1...\textmd{b}_k} \right)\nonumber\\
&&\left({U^{\textmd{a}_1...\textmd{a}_k}}\otimes\id ^{\textmd{b}_1...\textmd{b}_k}\right)\left(\rho_{00}^{\textmd{a}_1\textmd{b}_1}\otimes...\otimes \rho_{00}^{\textmd{a}_k\textmd{b}_k} \right) \left( {U^{\textmd{a}_1...\textmd{a}_k}}^\dagger \right.\nonumber\\
&&\left.  \otimes\id^{\textmd{b}_1...\textmd{b}_k} \right)\left({V_{m_1n_1}^{\textmd{a}_1\dagger}} \otimes...\otimes
{V_{m_kn_k}^{a_k\dagger}} \otimes\id^{b_1...b_k} \right),
\label{pi-m1n1-mknk}
\end{eqnarray}
where $U^{\textmd{a}_1...\textmd{a}_k}$ is an arbitrary unitary operator and  $V_{m_jn_j}^{\textmd{a}_j}$  are the operators in Eq. (\ref{vmn}).  For different states $\pi_{\{m_in_i\}}$,
\begin{eqnarray}
 \pi_{\{m_in_i\}} \pi_{ \{m^{\prime}_i n^{\prime}_i\}}=0,
 \label{orthognal-multi}
\end{eqnarray}
holds.

A proof for this Lemma is presented in the Appendix.  Using the
orthogonality property (\ref{orthognal-multi}), and the purity
 of the density operator $\pi_{\{m_in_i\}} $,  the channel output entropy can be written as
\begin{eqnarray}
&&S\Bigg(\Lambda^{\textmd{P}}_{\textmd{a}_1...\textmd{a}_k}\left(\left(U^{\textmd{a}_1...\textmd{a}_k}\otimes\id^{\textmd{b}_1...\textmd{b}_k} \right)
 \big(\rho^{\textmd{a}_1\textmd{b}_1}_{00} \otimes...\otimes \rho^{\textmd{a}_k\textmd{b}_k}_{00} \big)\right.\nonumber\\
&&\left. \big( {U^{\textmd{a}_1...\textmd{a}_k}}^\dagger \otimes\id^{\textmd{b}_1...\textmd{b}_k}
  \big) \right) \Bigg)
= S\left(\sum_{\{m_in_i\}} q_{\{m_in_i\}}\pi_{\{m_in_i\}}\right)\nonumber\\
&&= H\left(\{q_{\{m_in_i\}}\}\right),
\label{multi-u-ivariantt}
\end{eqnarray}
where $H\left(\{ p_{i}\}\right)
=-\sum_{i} p_{i} \log p_{i}$ is the Shannon entropy. Consequently, the channel output entropy  is just determined by the channel
 probabilities $q_{\{m_in_i\}}$ and it is invariant under
 unitary encoding. Therefore, both local encoding and global encoding leads to the same capacity in Eqs. (\ref{local-dc}) and (\ref{global-dc}). That is
\begin{subequations}
 \begin{eqnarray}
\label{c-bell-k-copy}
C^{k\textmd{-copy}\textmd{}}_\textmd{un,B}&=&\log d_{1}^2+\log d_{2}^2+...+\log d_{k}^2\nonumber\\
&-&H\left(\{q_{\{m_in_i\}}\}\right),\\
&&\nonumber\\
&\neq& k \hspace{1mm} C^{\textmd{one-copy}\textmd{}}_\textmd{un,B}.
\end{eqnarray} 
\end{subequations}
 The subscript \textquotedblleft \textmd{B}\textquotedblright \hspace{0.2mm} refers to a Bell state. As we can see from Eq. (\ref{c-bell-k-copy}), for a \emph{correlated} Pauli channel,  the capacity of  $k$ copies of a Bell state is not additive except when $\mu_{jl}=0$ for all $j$ and $l$, i.e. the case of an \emph{uncorrelated} Pauli channel with $ q_{\{m_in_i\}}=q_{m_1n_1}...q_{m_kn_k}$. Then the capacity for $k$ copies  is $k$ times  the capacity of a single copy with dimension $d^2$. That is 
 \begin{eqnarray}
C^{k\textmd{-copy,unco}\textmd{}}_\textmd{un,B}&=&k \left (\log d^2-H\left(\{q_{mn}\}\right) \right)\nonumber\\
&=& k \hspace{1mm}C^{\textmd{ one-copy,unco}\textmd{}}_\textmd{ un,B}.
\label{c-bell-k-copy-1}
\end{eqnarray} 
If $\mu_{jl}=1$ for all $j$ and $l$,  i.e. the case of a \emph{ fully correlated} Pauli channel with $ q_{\{m_in_i\}}=q_{mn}$, by using Eq. (\ref{c-bell-k-copy}), we have 
\begin{eqnarray}
C^{k\textmd{-copy}\textmd{}}_\textmd{un,B,f}&=&\log d_{}^2+...+\log d_{}^2-H\left(\{q_{mn}\}\right)\nonumber\\
&=&k\left(\log d_{}^2- \frac {H\left(\{q_{mn}\}\right)} {k} \right).
\label{c-bell-k-copy-d}
\end{eqnarray} 
Since $H\left(\{q_{mn}\}\right)$ is a constant value, in the limit of many copies $k$, by using Eq. (\ref{c-bell-k-copy-d}), we can reach the capacity $\log d^2$ per single copy. This is the highest capacity that we can reach for a  $d^2$ dimensional  system. 
\begin{center}
\begin{tabular}{c  }
\includegraphics[width=9cm]{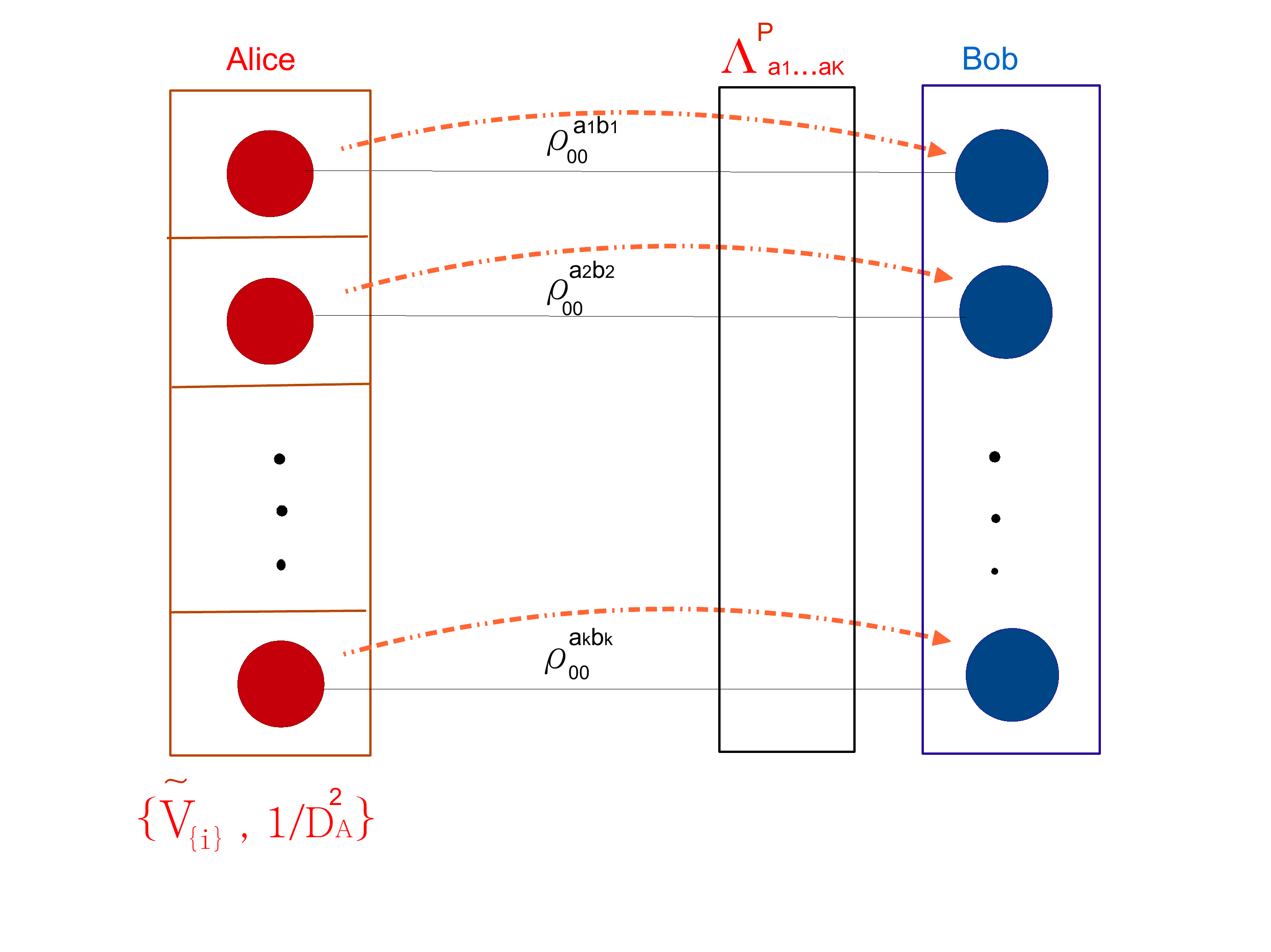}
\end{tabular}
\label{multi}
\end{center}
Fig. 2. (Color online) Multipartite super dense coding with $k$ copies of
 a  Bell state.  $\rho_{00}^{\textmd{a}_j\textmd{b}_j}$ is the $j$th copy   with   dimension $d_{j}^2$. The  dashed red curves show the transmission channels. Since the  $k$ channels can be correlated, the action of a
 global channel has been denoted by a \emph{correlated} Pauli channel
 $\Lambda_{\textmd{a}_1...\textmd{a}_k}^\textmd{P}$. There is no noise on  Bob's side. The optimal encoding for this case is the ensemble   $ \{\tilde{V}_{\{i\}},  \frac{1}{D_A ^2}\}$ (see main text).\\
\subsection{$k$ copies of a Bell diagonal state and a fully correlated Pauli channel \label{fully-Pauli}  }
Here, we give another example for which the capacity is exactly determined. This is the case of  $k$
copies of a Bell diagonal state and a \emph{fully correlated}  Pauli channel. 

As we defined in Sec. \ref{sec2-1},  when  $\mu_{jl}=1$ for all $j$ and $l$,  the channel is called a \emph{fully correlated} Pauli channel. For $d=2$, the operators  $V_{mn}$   are either the identity or the Pauli
operators $\sigma_m$, i.e.
\begin{eqnarray}
\sigma_0=\begin{pmatrix}
  1 & 0 \\
  0 & 1
\end{pmatrix},\hspace{1.5mm}
\sigma_1=\begin{pmatrix}
  0& 1 \\
  1& 0
\end{pmatrix},\hspace{1.5mm}\nonumber\\
\sigma_2=\begin{pmatrix}
  0 & -i \\
  i & 0 
  \end{pmatrix},\hspace{1.5mm}
\sigma_3=\begin{pmatrix}
  1 & 0 \\
  0 & -1
     \end{pmatrix}.
\end{eqnarray}
Thus, the channel, for  an arbitrary number of  parties, can be written as
\begin{eqnarray}
\Lambda  ^{\textmd{f}}(\xi)= \sum_m  q_m (\sigma_m \otimes...\otimes \sigma_m) \xi (\sigma_m \otimes...\otimes \sigma_m),
\label{fully-cor-channel-2}
\end{eqnarray}
where $ \sum_{m=0}^3 q_m = 1$. The superscript   \textquotedblleft \textmd{f}\textquotedblright \hspace{0.2mm} refers to a fully correlated Pauli channel. A Bell diagonal state is  a convex combination of the four Bell states. That is $ \rho_{Bd} = \sum^3_{n=0}  p_n\rho_n $, where $\rho_n$ is a Bell state, $p_n  \geq  0$,
and  $\sum^3_{n=0} p_n = 1$.  The subscript    \textquotedblleft \textmd{Bd}\textquotedblright \hspace{0.2mm} stands for a Bell
diagonal state. We here determine  both unitaries
$U^{\textmd{lo}}_{\textmd{min}}$ and  $U^{\textmd{g}}_{\textmd{min}}$. To do so, we first show that the von 
Neumann entropy of $k$ copies of a Bell diagonal state $\rho_{Bd}$  after applying an arbitrary unitary operator $U^{ \textmd{a}_1...  \textmd{a}_k}$  , and a \emph{fully correlated} Pauli channel (\ref{fully-cor-channel-2})  is lower bounded as
\begin{eqnarray}
&&S\bigg(\Lambda^{\textmd{f}}_{\textmd{a}_1...\textmd{a}_k\textmd{b}_1...\textmd{b}_k}\Big(\left(U^{\textmd{a}_1...\textmd{a}_k}\otimes\id^{\textmd{b}_1...\textmd{b}_k} \right)
 \left(\rho_{\textmd{Bd}}^{\textmd{a}_1\textmd{b}_1} \otimes...\otimes  \right. \nonumber\\
&& \left.  \left. \rho_{\textmd{Bd}}^{\textmd{a}_k\textmd{b}_k} \right) \big( {U^{\textmd{a}_1...\textmd{a}_k}}^\dagger \otimes\id^{\textmd{b}_1...\textmd{b}_k}
  \big) \right) \bigg)\nonumber\\
  &\geq&\sum_m  q_m S\bigg( (\sigma_m \otimes...\otimes \sigma_m) \Big(\left(U^{\textmd{a}_1...\textmd{a}_k}\otimes\id^{\textmd{b}_1...\textmd{b}_k} \right)
 \left(\rho_{\textmd{Bd}}^{\textmd{a}_1\textmd{b}_1}  \right. \nonumber\\
&& \left.  \left.  \otimes...\otimes \rho_{\textmd{Bd}}^{\textmd{a}_k\textmd{b}_k} \right) \big( {U^{\textmd{a}_1...\textmd{a}_k}}^\dagger \otimes\id^{\textmd{b}_1...\textmd{b}_k}
  \big) \right) (\sigma_m \otimes...\otimes \sigma_m) \bigg)\nonumber\\  
&=&S
 \left(\rho_{\textmd{Bd}}^{\textmd{a}_1\textmd{b}_1}\otimes...\otimes \rho_{\textmd{Bd}}^{\textmd{a}_k\textmd{b}_k} \right),
 \label{ fully-lowerbound-BD }
 \end{eqnarray}
where we used the concavity property of the von Neumann entropy. The lower bound in Eq. (\ref{ fully-lowerbound-BD }) is reachable by choosing  $U^  {\textmd{a}_1...\textmd{a}_k} = \id$ : 
\begin{eqnarray}
&&S\bigg(\Lambda^{\textmd{f}}_{\textmd{a}_1...\textmd{a}_k\textmd{b}_1...\textmd{b}_k}\Big(
 \rho_{\textmd{Bd}}^{\textmd{a}_1\textmd{b}_1} \otimes...\otimes 
     \rho_{\textmd{Bd}}^{\textmd{a}_k\textmd{b}_k}   \Big)\bigg)\nonumber\\
  &=& S\bigg(\sum_{n_1...n_k} p_{n_1}...p_{n_k}  \sum_m  q_m \underbrace{ (\sigma_m \otimes \sigma_m) 
  \rho_{n_1}(\sigma_m \otimes \sigma_m)} _{\rho_{n_1}}\nonumber\\  
&&  \otimes...\otimes \underbrace{(\sigma_m \otimes \sigma_m) 
  \rho_{n_k}(\sigma_m \otimes \sigma_m) }_{\rho_{n_k}} \bigg) \nonumber\\ 
&=&S  \left(\rho_{\textmd{Bd}}^{\textmd{a}_1\textmd{b}_1}\otimes...\otimes \rho_{\textmd{Bd}}^{\textmd{a}_k\textmd{b}_k} \right).
\label{}
\end{eqnarray}
As  $U^  {\textmd{a}_1...\textmd{a}_k} = \id $, the super dense coding capacities with both local encoding and global encoding (\ref{local-dc}) and (\ref{global-dc}) are the same. Therefore, Alice cannot do better encoding than local encoding on each copy of the Bell diagonal states. According to Eq. (\ref{local-dc}), the capacity of $k$  copies of a Bell diagonal state, when the states are sent through a \emph{fully correlated} Pauli channel (\ref{fully-cor-channel-2}), is additive, i.e. 
\begin{eqnarray}
C^{k\textmd{-copy}}_\textmd{un,Bd,f}&=&k\big(2-S( \rho_{\textmd{Bd}})\big)\nonumber\\
&=& k \hspace{1mm}C^{\textmd{one-copy}}_\textmd{un,Bd,f} .
\label{capacity-BD}
\end{eqnarray} 
The capacity (\ref{capacity-BD}) shows that for fully correlated channels no information at all is lost to the environment and this class of channels behaves like a noiseless one. 

For k copies of a Bell state, by using Eq. (\ref {capacity-BD}), and the  purity of a Bell state, we have
\begin{eqnarray}
C^{k\textmd{-copy}}_\textmd{un,B,f}&=&2k ,
\label{k-copy-bell}
\end{eqnarray} 
which is the highest amount of information transfer for $2k$ parties where each of them has  a two-level system.

\subsection{GHZ state and a fully correlated Pauli channel}

Another example for which we can determine  both unitaries $U_{\textmd{min}}^{\textmd{lo}}$ and $U_{\textmd{min}}^{\textmd{g}}$, is a $\ket{GHZ}$ state of 2-dimensional subsystems distributed between $2k - 1$ Alices and a single Bob. The channel here is a \emph{fully correlated} Pauli channel, as defined via Eq. (\ref{fully-cor-channel-2}).  For a system of $2k$ parties, the $\ket{GHZ}$ state can be written as
\begin{eqnarray}
\ket{GHZ}_{2k}=\frac{1}{\sqrt{2}}\sum_0^1  \ket {j^{(1)}...j^{(2k)}}.
\end{eqnarray} 
Since the minimum value of a von Neumann entropy is
zero, and since a $ \ket{GHZ}$ state is invariant under the action of a \emph{ fully correlated} Pauli channel, we have
\begin{eqnarray}
&&S\bigg(\Lambda^{\textmd{f}}_{\textmd{a}_1...\textmd{a}_{2k-1}\textmd{b}}\Big( \ket{GHZ}_{2k}\bra{GHZ}    \Big)\bigg)\nonumber\\
    &=&S\bigg(\sum_m  q_m \underbrace{ (\sigma_m \otimes...\otimes \sigma_m) \Big(   \ket{GHZ}_{2k}\bra{GHZ}   \Big)} \nonumber\\
&&  \underbrace{ (\sigma_m \otimes...\otimes \sigma_m) \bigg)}_{\ket{GHZ}_{2k}\bra{GHZ} }\nonumber\\  
&&\nonumber\\
&=&S  \left(  \ket{GHZ}_{2k}\bra{GHZ} \right)\nonumber\\
&=&0.
\label{}
\end{eqnarray}
Therefore, by using $U_{\textmd{min}}^{\textmd{lo}}= U_{\textmd{min}}^{\textmd{g}}=\id$, we  can  reach the zero entropy. Then, the super dense coding capacity,  according to Eq. (\ref{local-dc}), reads
\begin{eqnarray}
C^{\textmd{f}}_\textmd{un,GHZ}=2k.
\label{k-copy-ghz}
\end{eqnarray} 
Here, the \emph{fully correlated}  Pauli channel, for a $\ket{GHZ}$ state, behaves like a noiseless channel and again no information is lost through the channel.

\subsection{$k$ copies of an arbitrary state and  an uncorrelated depolarizing channel}

The last example for which we determine  the capacity exactly  is the case of  the $k$
copies of an arbitrary state $\rho^{\textmd {ab}}$, where the diminetion of both $\rho^{\textmd a}$ and  $\rho^{\textmd b}$ is $d$, in the presence of an \emph{uncorrelated} depolarizing  channel.   

A \emph{d}-dimensional depolarizing channel is a channel that transmits a  quantum system intact with the probability $1-p$	and	randomizes its state with	the  probability $p$. This channel is a special case of a \emph{d}-dimensional Pauli channel with the probability parameters
\begin{equation}
q_{mn}=\left\{ 
\begin{array}{lcl}
1-p+\frac {p}{d^2} &,&  m=n=0 \\
 \frac{p}{d^2} &,&       \mbox{otherwise}.
\end{array}
 \right.
\end{equation}
 with $0\leq p \leq 1$, and $m,n=0,...,d-1$. 

In \cite{zahra-paper}, for a bipartite state,  we showed that the von Neumann entropy of
a state that was sent through the depolarizing channel with \emph{uncorrelated} noise is independent of any local unitary transformations that were performed before the action of the channel, i.e.
\begin{eqnarray}
&&S\bigg(\Lambda_{{\textmd{a}\textmd{b}}}^{\textmd{dep}}\Big(  \big(U \otimes\id^{\textmd{b}}\big)   \rho^{\textmd{a}\textmd{b}}  \big(U^{ \dagger}  \otimes\id^{\textmd{b}} \big) \Big)\bigg)=S\left(\Lambda_{{\textmd{a}\textmd{b}}}^{\textmd{dep}}  \left( \rho^{\textmd{a}\textmd{b}}\right)  \right).\nonumber\\
\label{dep-bipartite}
\end{eqnarray}
In this section,  we show that the same result is valid for  a given  resource state  $\rho^{\textmd{a}_1\textmd{b}_1}\otimes...\otimes\rho^{\textmd{a}_k\textmd{b}_k}$, and  the \emph{uncorrelated}  depolarizing channel $ \Lambda_{{\textmd{a}_1...\textmd{a}_k\textmd{b}_1...\textmd{b}_k}}^{\textmd{dep}}$.  A proof for this statement  is as follows.
\begin{eqnarray}
&&S\Bigg(\Lambda_{{\textmd{a}_1...\textmd{a}_k\textmd{b}_1...\textmd{b}_k}}^{\textmd{dep}}\Big(  \big(U_{\textmd{min}}^{\textmd{lo}} \otimes\id^{\textmd{b}_1... \textmd{b}_k}\big) \nonumber\\
&&\left. \left( \rho^{\textmd{a}_1\textmd{b}_1}\otimes...\otimes\rho^{\textmd{a}_k\textmd{b}_k}\right)  \big(U_{\textmd{min}}^{\textmd{lo}\dagger} \otimes\id^{\textmd{b}_1... \textmd{b}_k} \big) \right) \Bigg)\nonumber\\
&=&S\Bigg( \Lambda_{{\textmd{a}_1\textmd{b}_1}}^{\textmd{dep}} \left(  \big(U_\textmd{min}^{ \textmd{a}_1}  \otimes\id^{\textmd{b}_1}\big) \rho^{\textmd{a}_1\textmd{b}_1}\big(U_\textmd{min}^{ \textmd{a}_1\dagger}  \otimes\id^{\textmd{b}_1}\big)\right) \otimes...\otimes\nonumber\\
&&\Lambda_{{\textmd{a}_k\textmd{b}_k}}^{\textmd{dep}} \left(  \big(U_\textmd{min}^{ \textmd{a}_k}  \otimes\id^{\textmd{b}_k}\big) \rho^{\textmd{a}_k\textmd{b}_k}\big(U_\textmd{min}^{ \textmd{a}_k\dagger}  \otimes\id^{\textmd{b}_k}\big)\right)\Bigg) \nonumber\\
&&= S\left(\Lambda_{{\textmd{a}_1\textmd{b}_1}}^{\textmd{dep}}  \left( \rho^{\textmd{a}_1\textmd{b}_1}\right)  \right) +...+S\left(\Lambda_{{\textmd{a}_k\textmd{b}_k}}^{\textmd{dep}}  \left( \rho^{\textmd{a}_k\textmd{b}_k}\right)  \right),     
\label{2-dep}
\end{eqnarray}
where in the last equality we used Eq. (\ref{dep-bipartite}), and  the additivity of the von Neumann entropy.  This proves our above claim. Therefore, by using (\ref{2-dep}),  and according to Eq. (\ref{local-dc}), the super dense coding capacity, for  $k$ copies  of a resource state $\rho^{\textmd{a}\textmd{b}}$, is given by
\begin{eqnarray}
C^{\textmd{k-copy}}_{\textmd{un,dep}}&=&k\bigg(\log d+ S\left(\Lambda_{\textmd{b}}^\textmd{dep} \left(
\rho_{\textmd{b}}\right)\right)-S\left(\Lambda_{{\textmd{a}\textmd{b}}}^{\textmd{dep}}  \left( \rho^{\textmd{a}\textmd{b}}\right)  \right)\bigg)\nonumber\\
&=&k \hspace{1mm}C^{\textmd{one-copy}}_{\textmd{un,dep}}.
\label{2}
\end{eqnarray}
In Table 1, the above mentioned  examples  and also some of the unsolved examples  are summarized.  
\begin{widetext}  

\begin{center}
\begin{tabular}{ |  c ||    c  |    c  | c l |l|}\hline 

 & {\footnotesize  Correlated Pauli channel } & {\footnotesize  Fully correlated Pauli channel } &{\footnotesize Uncorrelated depolarizing channel}  & \\ 

 \backslashbox{ {\footnotesize Resource state} }{  {\footnotesize Channel} }  &   {\footnotesize only on the Alices' sides}& {\footnotesize $  (\mu_{jl}=1)$}  {\footnotesize and ($d=2$)} &{\footnotesize  $(\mu_{jl}=0)$ (arbitrary dimension)} & \\ 
  
 &   {\footnotesize  (arbitrary dimension) }&  &  & \\ \hline\hline
{\footnotesize $\mathrm{k}$ copies of a  Bell state } & {\footnotesize Eq. (\ref{c-bell-k-copy})}   &    {\footnotesize Eq. (\ref{k-copy-bell})}&  {\footnotesize  Eq. (\ref{2})} &  \\ \hline

 {\footnotesize $\mathrm{k}$ copies of a Bell diagonal state } &\footnotesize 
{\color{red} open } & {\footnotesize Eq. (\ref{capacity-BD})} & {\footnotesize  Eq. (\ref{2})}   & \\ \hline

 {\footnotesize GHZ state with 2k parties  } &  \footnotesize  {\color{red} open }  & { \footnotesize Eq. (\ref{k-copy-ghz})}&   \footnotesize  {\color{red} open }  & \\ \hline 
{\footnotesize  $\mathrm{k}$ copies of an arbitrary state $\rho^{\textmd ab}$} &   \footnotesize {\color{red} open } &  \footnotesize  {\color{red} open }  &{\footnotesize  Eq. (\ref{2})} & \\ \hline

    \end{tabular}
\end{center}

\vspace{2mm}

Table 1.   A summary of the solved examples of  multipartite resource states and channels for super dense coding. Here, some of the unsolved examples are also mentioned. 
\end{widetext}

\section{distributed super dense CODING WITH  NON-UNITARY ENCODING}

In a multipartite super dense coding scheme with non-unitary encoding,  
instead of the unitary operators  $W_{\{i\}}^{\textmd {a}_1...\textmd {a}_k}$
considered   
in the previous  sections, the  Alices apply the CPTP maps $\Gamma_{\{ i\}}^ {\textmd {a}_1...\textmd {a}_k}$ on their side of the shared state $\rho^{\textmd {a}_1...\textmd {a}_k \textmd {b}} $ and thereby perform the encoding  via the states $\rho_{\{ i\}}= (\Gamma_{\{ i\}} ^ {\textmd {a}_1...\textmd {a}_k}\otimes\id^ {\textmd{b}}) \left(\rho^{\textmd {a}_1...\textmd {a}_k \textmd {b}}\right) $. The rest of the scheme is similar to the case of unitary encoding. The Alices send the encoded state $\rho_{\{ i\}}$, with the probability $p_{\{i\}}$,  through the 
covariant channel  to Bob. The super dense coding capacity  is then 
the maximum of the Holevo quantity with respect to the CPTP maps $\Gamma_{\{ i\}} ^ {\textmd {a}_1...\textmd {a}_k}$ and  the probabilities $p_{\{i\}}$.  
We first consider the case for which the Alices are again restricted to  
local CPTP maps $\Gamma_{ i_j}^ {\textmd {a}_j}$.  
For this situation, the optimization of the Holevo quantity in the presence 
of a covariant channel is given in the following lemma.

\textbf{Lemma 3.} Let 

\begin{eqnarray}
\chi_{\textmd {non-un}}^{\textmd {lo}} &=&S\bigg (
\sum_{\{i\}} p_{\{i\}}
 \Lambda_{\textmd {a}_1...\textmd {a}_k\textmd {b}}^\textmd {c}\left(\rho_{\{i\}}  \right)\bigg)\nonumber\\
&-&\sum_{\{i\}} p_{\{i\}} S\bigg(\Lambda_{\textmd {a}_1...\textmd {a}_k\textmd {b}}^\textmd {c} \left(
\rho_{\{i\}}\right)\bigg),
 \label{Holevo-non-unitary}
\end{eqnarray}
be the Holevo quantity with  
\begin{eqnarray}
\rho_ {\{i\}}=\Big(  \Gamma_{i_1}^{\textmd {a}_1}\otimes \Gamma_{i_2}^{a_2}\otimes...\otimes  \Gamma_{i_k}^{\textmd {a}_k}   \otimes \id^\textmd {b}\Big) \left (\rho^{\textmd {a}_1...\textmd {a}_k\textmd {b}}\right),
\end{eqnarray}
 and $\Lambda^{\textmd c}_{\textmd {a}_1...\textmd {a}_k\textmd {b}}$ be a
covariant channel.  Let 
\begin{eqnarray}
\Gamma_\textmd{min}^{\textmd{lo}}:=\Gamma_{\textmd {min}}^{\textmd {a}_1}\otimes \Gamma_{\textmd {min}}^{a_2}\otimes...\otimes  \Gamma_{\textmd {min}}^{\textmd {a}_k}
\label{3}
\end{eqnarray}
be the map that minimizes the
von Neumann entropy after application of this map and
the covariant channel  to the initial state $ \rho^{\textmd {a}_1...\textmd {a}_k\textmd {b}}$. 
 Then the super dense coding capacity  $C_{\textmd{non-un}}^{\textmd{lo}}$ is given by
\begin{eqnarray}
&C_{\textmd{non-un}}^{\textmd{lo}}=&\log D_\textmd{A}+ S\left(\Lambda_{\textmd{b}}^\textmd{} \left(
\rho_{\textmd{b}}\right)\right)\nonumber\\
&& -S\Big(\Lambda_{\textmd{a}_1...\textmd{a}_k\textmd{b}}^\textmd{c} \big(   \Gamma_\textmd{min}^{\textmd{lo}} (\rho^{\textmd{a}_1...\textmd{a}_k\textmd{b}}) \big)\Big),
\label{1-local-non-unitary-commutative}
\end{eqnarray}
where    $\mathrm{\tr}_{\textmd{a}_1...\textmd{a}_k}  \Lambda_{\textmd{a}_1...\textmd{a}_k\textmd{b}}^\textmd{c} \left(\rho^{\textmd{a}_1...\textmd{a}_k\textmd{b}}\right)=\Lambda_{\mathrm{b}}^\textmd{} \left(
\rho_{\mathrm{b}}\right)$ and $D_\textmd{A}=d_{\textmd{a}_1} d_{\textmd{a}_2}...d_{\textmd{a}_k}$.\\
A proof for this Lemma is shown in the Appendix.

Now,  if  the Alices are allowed to perform  global operations, with an argument similar to Lemma 3,  we can show that the super dense coding capacity  is  
\begin{eqnarray}
&C_{\textmd{non-un}}^\textmd{g}=& \log D_\textmd{A}+ S\left(\Lambda_{\textmd{b}}^\textmd{} \left(
\rho_{\textmd{b}}\right)\right) \nonumber\\
& &-S\bigg(\Lambda_{\textmd{a}_1...\textmd{a}_k\textmd{b}}^\textmd{c}    \Big( \Gamma_\textmd{min}^{  \textmd{g}  }  (\rho^{\textmd{a}_1...\textmd{a}_k\textmd{b}}) \Big)\bigg).
\label{global-non-unitary-commutative}
\end{eqnarray}
 Here,  $\Gamma_\textmd{min}^{\textmd{g}}$  is a pre-processing that the  Alices globally perform  on the initial state   $ \rho^{\textmd{a}_1...\textmd{a}_k\textmd{b}}$   before applying the optimal local  unitary operators $\tilde{V}_{\{i\}}$.  The pre-processing   $\Gamma_\textmd{min}^{\textmd{g}}$  minimizes the output von Neumann entropy after applying it and the channel to the initial state.  The capacity  (\ref{global-non-unitary-commutative})  is reachable by the optimal ensemble    $\Big\{\tilde{\Gamma}_{\{i\}}(\xi)=(\tilde{V}_{\{i\}} \otimes\id^\textmd{b})
\left[\Gamma_\textmd{min}^{\textmd{g}} (\xi)\right]
(\tilde{V}_{\{i\}}^\dagger \otimes\id^\textmd{b}),\tilde{p}_{\{i\}}=\frac{1}{D^2_{\textmd{A}}}\Big\}.$ 

Similar to unitary encoding, the capacities (\ref{1-local-non-unitary-commutative}) and (\ref{global-non-unitary-commutative}) can be also written for the special case of a  Pauli channel. 
\section{conclusion}
In summary, we discussed in this paper the multipartite super dense coding  
scenario of  many senders 
and a single receiver in the presence of a covariant  channel.  
Considering (non)unitary encoding,  for both cases of local and global 
encoding, and up to some pre-processing on the resource state,   we found  
expressions for the capacity.    In general, the pre-processing  is not  
determined  and  it is  an open question. For unitary encoding,  we 
found examples for  which the pre-processing can be determined and turns 
out to be the identity operator. For the mentioned examples,  Alices cannot 
do better  than local encoding.  We also showed that for some of these  
examples  Alices cannot do better than unitary encoding. 
 
These results can be seen as first steps in several directions of
future  research. For example,  it would be  interesting to consider other 
types of channels rather than  Pauli channels and also other types of 
memories.   
It would be also interesting to consider  the case  where we have more than 
one receiver.  The case of two receivers, for  noiseless channels,  is 
discussed in \cite{Dagmar} where
 some of the Alices send their information to the first Bob while the others 
send theirs  to the second Bob. 
 The two receivers are restricted to perform local operations and classical 
communication among themselves. To the best of our knowledge, for 
this situation the  exact super dense  coding capacity is still an open 
question even for  noiseless channels.     
For bipartite super dense coding, we showed  previously that there are examples for 
which the non-unitary encoding leads to a  better capacity  than unitary one. 
It is still an open problem to establish whether this can happen also
in the multipartite case.
 
{\bf Acknowledgments}: This work was partially supported by   Deutsche Forschungsgemeinschaft (DFG) and the EU-project CORNER.

\newpage

\section{APPENDIX}
.\newline \textbf{Proof for Lemma 2:} To prove the Lemma, we first show
that the statement
\begin{eqnarray}
&&\bra{\Phi_{00}^{\textmd{a}_1b_1}...\Phi_{00}^{\textmd{a}_k\textmd{b}_k}} ({U^{\textmd{a}_1...\textmd{a}_k}}^\dagger)
\left({V_{m_1n_1}^{\textmd{a}_1\dagger}} V_{m^{\prime}_1 n^{\prime}_1 }\otimes...\otimes{V_{m_kn_k}^{\textmd{a}_k\dagger}} 
\right.\nonumber\\
&&  \left.V_{m^{\prime}_kn^{\prime}_k} \otimes \id^{\textmd{b}_1...\textmd{b}_k}\right)\left({U^{\textmd{a}_1...\textmd{a}_k}}\right)\ket{\Phi_{00}^{\textmd{a}_1\textmd{b}_1}... \Phi_{00}^{\textmd{a}_k\textmd{b}_k}}=0,
\label{pre-proof-orthogonal}
\end{eqnarray}
holds. 
By using the definition of a Bell state $\ket{\Phi_{00}}= \frac{1}{\sqrt d}\sum_{j=0}^{d-1}\ket{jj}$ for a Bell state, we have

\begin{eqnarray}
&&\bra{\Phi_{00}^{\textmd{a}_1\textmd{b}_1}...\Phi_{00}^{\textmd{a}_k\textmd{b}_k}} ({U^{\textmd{a}_1...\textmd{a}_k}}^\dagger)
\left({V_{m_1n_1}^{\textmd{a}_1\dagger}} V_{{m}_1^{\prime}{n}_1^{\prime}}^{\textmd{a}_1}\otimes...\otimes
\right.\nonumber\\
&&\left.{V_{m_kn_k}^{\textmd{a}_k\dagger}} V_{{m}_k^{\prime}{n}_k^{\prime}}^{\textmd{a}_k}\otimes \id^{\textmd{b}_1...\textmd{b}_k}\right) \left({U^{\textmd{a}_1...\textmd{a}_k}}\right) \ket{\Phi_{00}^{\textmd{a}_1\textmd{b}_1}... \Phi_{00}^{\textmd{a}_k\textmd{b}_k}}\nonumber\\
&=&\sum_{j_1...j_k}
\sum_{{j}_1^{\prime}...{j}_k^{\prime}}\bra{j_1j_1...j_kj_k}({U^{\textmd{a}_1...\textmd{a}_k}}^\dagger)\left({V_{m_1n_1}^{\textmd{a}_1\dagger}} V_{{m}_1^{\prime}{n}_1^{\prime}}^{\textmd{a}_1} \right.\nonumber\\
&& \left.\otimes...\otimes {V_{m_kn_k}^{\textmd{a}_k\dagger}} V_{{m}_k^{\prime}{n}_k^{\prime}}^{\textmd{a}_k}
 \otimes \id^{\textmd{b}_1...\textmd{b}_k}\right) \left({U^{\textmd{a}_1...\textmd{a}_k}}\right)\ket{{j}_1^{\prime} {j}_1^{\prime}...{j}_k^{\prime}{j}_k^{\prime}}  \nonumber\\
&=&\sum_{j_1...j_k}\bra{j_1...j_k} {U^{\textmd{a}_1...\textmd{a}_k}}^\dagger
 \left({V_{m_1n_1}^{\textmd{a}_1\dagger}} V_{{m}_1^{\prime}{n}_1^{\prime}}^{\textmd{a}_1}\otimes...\otimes\right.
\nonumber\\
&&\left. {V_{m_kn_k}^{\textmd{a}_k\dagger}} V_{{m}_k^{\prime}{n}_k^{\prime}}^{\textmd{a}_k} \right ){U^{\textmd{a}_1...\textmd{a}_k}}\ket{j_1...j_k}\nonumber\\
&=& \mathrm{\tr}_{\textmd{a}_1...\textmd{a}_k}\left( {V_{m_1n_1}^{\textmd{a}_1\dagger}} V_{{m}_1^{\prime}{n}_1^{\prime}}^{\textmd{a}_1}\otimes...\otimes
{V_{m_kn_k}^{\textmd{a}_k\dagger}}
V_{{m}_k^{\prime}{n}_k^{\prime}}^{\textmd{a}_k} \right)\nonumber\\
&=& \delta_{m_1{m}_1^{\prime}} \delta_{n_1{n}_1^{\prime}}...\delta_{m_k{m}_k^{\prime}}\delta_{n_k{n}_k^{\prime}},
\label{1358}
\end{eqnarray}
where in the last line we have used $\mathrm{\tr} V_{mn} V_{{m}^{\prime}{n}^{\prime}}^\dagger =
d\delta_{m{m}^{\prime}} \delta_{n{n}^{\prime}}$. Different
states $\pi_{\{m_in_i\}}$ have at least one different index
for $m_i$ or $n_i$. Then by using Eq. (\ref{1358}), the statement of equation 
(\ref{pre-proof-orthogonal}) is proved. Subsequently, we arrive at
\begin{eqnarray}
&&( \rho_{00}^{\textmd{a}_1\textmd{b}_1}\otimes...\otimes \rho_{00}^{\textmd{a}_k\textmd{b}_k} )
( {U^{\textmd{a}_1...\textmd{a}_k}}^\dagger)\left({V_{m_1n_1}^{\textmd{a}_1\dagger}} V_{{m}_1^{\prime}{n}_1^{\prime}}^{\textmd{a}_1}\otimes...\otimes V_{m_kn_k}^{\textmd{a}_k\dagger} \right.\nonumber\\
&&\left. V_{{m}_k^{\prime}{n}_k^{\prime}}^{\textmd{a}_k}\otimes \id^{\textmd{b}_1...\textmd{b}_k}\right)\left({U^{\textmd{a}_1...\textmd{a}_k}}\right)(\rho_{00}^{\textmd{a}_1\textmd{b}_1}\otimes...\otimes \rho_{00}^{\textmd{a}_k\textmd{b}_k}) =0.
\label{pre-proof-orthogona2}
\end{eqnarray}
By using Eq. (\ref{pre-proof-orthogona2}), for
$\pi_{\{m_in_i\}}\pi_{  \{m ^{\prime}_i n^{\prime}_i\}}$
we have
\begin{eqnarray}
&&\pi_{\{m_in_i\}}\pi_{  \{m ^{\prime}_i n^{\prime}_i\}}\nonumber\\
&=&\left(V_{m_1n_1}^{\textmd{a}_1} \otimes ... \otimes V_{m_kn_k}^{\textmd{a}_k} \otimes\id^{\textmd{b}_1...\textmd{b}_k} \right)
\left({U^{\textmd{a}_1...\textmd{a}_k}}\right)\underbrace{\rho_{00}^{\textmd{a}_1\textmd{b}_1}\otimes...\otimes  }\nonumber\\
&&\nonumber\\
&& \underbrace{
\rho_{00}^{\textmd{a}_k\textmd{b}_k}( {U^{\textmd{a}_1...\textmd{a}_k}}^\dagger)\left({V_{m_1n_1}^{\textmd{a}_1\dagger}} V_{{m}_1^{\prime}{n}_1^{\prime}}^{\textmd{a}_1}\otimes...\otimes
{V_{m_kn_k}^{\textmd{a}_k\dagger}} V_{{m}_k^{\prime}{n}_k^{\prime}}^{\textmd{a}_k}\otimes \right.}\nonumber\\
&&\nonumber\\
&&\underbrace{ \id^{\textmd{b}_1...\textmd{b}_k}\big)\left({U^{\textmd{a}_1...\textmd{a}_k}}\right)\rho_{00}^{\textmd{a}_1\textmd{b}_1}\otimes...\otimes \rho_{00}^{\textmd{a}_k\textmd{b}_k} }_{= 0}( {U^{\textmd{a}_1...\textmd{a}_k}}^\dagger )\nonumber\\
&&\nonumber\\
&&\left(V_{{m}_1^{\prime}{n}_1^{\prime}}^{\textmd{a}_1} 
 \otimes ... \otimes V_{{m}_k^{\prime}{n}_k^{\prime}}^{\textmd{a}_k} \otimes\id^{\textmd{b}_1...\textmd{b}_k} \right)\nonumber\\
&=&0 \hspace{0.9mm}, \nonumber
\end{eqnarray}
which completes the proof.
\hfill{$\Box$}\\

\textbf{Proof for Lemma 3:}  With an argument similar to Lemma 1, we first 
introduce an upper bound on the Holevo quantity (\ref{Holevo-non-unitary}), 
and in the next step, we show that the bound is attainable. By using the 
subadditivity of the von Neumann entropy, noting that the maximum entropy of 
a $d$-dimensional system is $\log d$, and since the map 
$\Gamma_\textmd{min}^{\textmd{lo}}$ gives the minimum output entropy, 
we have the upper bound

\begin{eqnarray}
&\chi_{\textmd {non-un}}^{\textmd {lo}}\leq& \log D_\textmd{A}+ S\left(\Lambda_{\textmd{b}}^\textmd{} \left(
\rho_{\textmd{b}}\right)\right)\nonumber\\
&& - S\Big(\Lambda_{\textmd{a}_1...\textmd{a}_k\textmd{b}}^\textmd{c} \big(\Gamma_\textmd{min}^{\textmd{lo}}
  (\rho^{\textmd{a}_1...\textmd{a}_k\textmd{b}}) \hspace{1mm} \big) \Big). 
  \label{non-unitary-upperbound}
\end{eqnarray}
The above bound is reachable by the ensemble  $\Big\{\tilde{\Gamma}_{\{i\}}(\xi)=(\tilde{V}_{\{i\}} \otimes\id^\textmd{b})
\left[\Gamma_\textmd{min}^{\textmd{lo}} (\xi)\right]
(\tilde{V}_{\{i\}}^\dagger \otimes\id^\textmd{b}),\tilde{p}_{\{i\}}=\frac{1}{D^2_{\textmd{A}}}\Big\} 
$.  In other words, the optimal encoding consists of a fixed pre-processing  
$\Gamma_\textmd{min}^{\textmd{lo}}$  (\ref{3})  and a subsequent unitary 
encoding $\tilde{V}_{\{i\}}$. 
We recognize  $\tilde{\chi}_{\textmd {non-un}}^{\textmd {lo}}$  as the Holevo quantity for the ensemble $\{\tilde{\Gamma}_{\{i\}}(\xi), \tilde{p}_{\{i\}} \} 
$ which is given by

\begin{eqnarray}
&&\tilde{\chi}_{\textmd {non-un}}^{\textmd {lo}}= S \bigg( \sum_{\{i\}}\frac{1}{D_\textmd A^2}\Lambda^{ \textmd c}_{\textmd{a}_1...\textmd{a}_k\textmd{b}}\left( \tilde{\Gamma}_{\{i\}}(\rho^{\textmd{a}_1...\textmd{a}_k\textmd{b}} ) \right) \bigg)\nonumber\\
&&-\sum_{\{i\}} \frac{1}{D_\textmd A^2}
S\bigg(    \Lambda^{ \textmd c}_{\textmd{a}_1...\textmd{a}_k\textmd{b}}\left( \tilde{\Gamma}_{\{i\}}(\rho^{\textmd{a}_1...\textmd{a}_k\textmd{b}} ) \right)    \bigg).
\label{optiensemble-NON-unitary}
\end{eqnarray}
 In the following, we show that the  above quantity (\ref{optiensemble-NON-unitary}) is equal to the bound on Eq. (\ref{non-unitary-upperbound}). 
 
The pre-processing $\Gamma_\textmd{min}^{\textmd{lo}}$  maps the  quantum 
state $ \rho^{\textmd{a}_1...\textmd{a}_k\textmd{b}}$ to another quantum 
state  $\rho^{ \prime\textmd{a}_1...\textmd{a}_k\textmd{b}}$. 
By using the decomposition  (\ref{decomposition-multi}) for  
$\Lambda^{ \textmd c}_{\textmd{a}_1...\textmd{a}_k\textmd{b}}
\left( \rho^{ \prime\textmd{a}_1...\textmd{a}_k\textmd{b}}\right)$,  
Eq. (\ref{average-lambda}), and the covariance 
property of the channel,  we find that 
the first term on the RHS of (\ref{optiensemble-NON-unitary}) is given
by
\begin{eqnarray}
&& \sum_{\{i\}}\frac{1}{D_\textmd A^2}\Lambda^{ \textmd c}_{\textmd{a}_1...\textmd{a}_k\textmd{b}}\left(
(\tilde{V}_{\{i\}}\otimes\id^{\textmd b})\rho^{\prime\textmd{a}_1...\textmd{a}_k\textmd{b}}({\tilde{V}_{\{i\}}}^{\dagger}\otimes\id^{\textmd b})
 \right)\nonumber\\
 &=& \sum_{\{i\}}\frac{1}{D_\textmd A^2}( \tilde{V}_{\{i\}}\otimes\id^{\textmd b})\Lambda^{ \textmd c}_{\textmd{a}_1...\textmd{a}_k\textmd{b}} 
(\rho^{\prime\textmd{a}_1...\textmd{a}_k\textmd{b}})({\tilde{V}_{\{i\}}}^{\dagger}\otimes\id^{\textmd b} 
 )\nonumber\\
&=&\frac{\id^{\textmd{a}_1...\textmd{a}_k}}{D_\textmd{A}}\otimes \Lambda_\textmd{b}( \rho_\textmd{b}) .
\label{average-state2}
\end{eqnarray}

Using the covariance property of the channel and  the invariance of the von Neumann entropy under unitary transformations,  the second term on the RHS of Eq. (\ref{optiensemble-NON-unitary}) can be written as

\begin{eqnarray}
&&\sum_{\{i\}} \frac{1}{D_\textmd A^2}
S\left(\Lambda^{\textmd c}_{\textmd{a}_1...\textmd{a}_k\textmd{b}}\left(
(\tilde{V}_{\{i\}}\otimes\id^\textmd{b})\rho^{\prime\textmd{a}_1...\textmd{a}_k\textmd{b}}({\tilde{V}_{\{i\}}}^{\dagger}\otimes\id^\textmd{b}) \right)\right)\nonumber\\
&=&\frac{1}{D_\textmd A^2} \sum_{{\{i\}}} S\Big((\tilde{V}_{{\{i\}}}\otimes \id^\textmd{b} )
\Big[\Lambda_{\textmd{a}_1...\textmd{a}_k\textmd{b}}^\textmd{c} (\rho^{\prime\textmd{a}_1...\textmd{a}_k\textmd{b}}) \Big]
(\tilde{V}_{{\{i\}}}^\dagger\otimes \id^\textmd{b})\Big)\nonumber\\
&=&S\bigg(\Lambda_{\textmd{a}_1...\textmd{a}_k\textmd{b}}^\textmd{c} \Big (\Gamma_\textmd{min}^{\textmd{lo}}  (\rho^{\textmd{a}_1...\textmd{a}_k\textmd{b}}) \hspace{1mm}  \Big) \bigg).
\label{average-entropy2}
\end{eqnarray}
Inserting Eqs. (\ref{average-state2}) and (\ref{average-entropy2}) into
Eq. (\ref{optiensemble-NON-unitary}), one finds that the Holevo quantity
$\tilde{\chi}_{\textmd {non-un}}^{\textmd {lo}}$ is equal to  the upper bound given in
Eq. (\ref{non-unitary-upperbound}) and consequently, this is the super dense
coding capacity. \hfill {$\Box$}


\end{document}